# Feeder bus service design under spatially heterogeneous demand

Li Zhen [a], Weihua Gu [a*]

[c] Department of Electrical Engineering, The Hong Kong Polytechnic University, Hong Kong SAR, China

**Abstract**

In rapidly sprawling urban areas and booming intercity express rail networks, efficiently designed feeder bus systems are more essential than ever to transport passengers to and from trunk-line rail terminals. When the feeder service region is sufficiently large, the spatial heterogeneity in demand distribution must be considered. This paper develops continuous approximation models for optimizing a heterogeneous fixed-route feeder network in a rectangular service region next to a rail terminal. Our work enhances previous studies by: (i) optimizing heterogeneous stop spacings along with line spacings and headways; (ii) accounting for passenger boarding and alighting numbers on bus dwell times and patron transfer delays at the rail terminal; and (iii) examining the advantages of asymmetric coordination between trunk and feeder schedules in both service directions. To tackle the increased modeling complexity, we introduce a semi-analytical method that combines analytically derived properties of the optimal solution with an iterative search algorithm. Local transit agencies can readily utilize this approach to design a real fixed-route feeder system.

This paper reveals many findings and insights not previously reported. For instance, integrating the heterogeneous stop spacing optimization further reduces the system cost (by 4% under specific operating conditions). The cost savings increase with demand heterogeneity but decrease with the demand rate and service region size. Choosing the layout of feeder lines where buses pick up and drop off passengers along the service region's shorter side also significantly lowers the system cost (by 6% when the service region's aspect ratio is 1 to 2). Furthermore, coordinating trunk and feeder schedules in both service directions yields an additional cost saving of up to 20%.

## 1. Introduction

The global trend of urbanization has accelerated in recent decades (Mor et al., 2020), leading to the growing urban sprawl and the emergence of new mega-cities (Yu et al., 2019). In response, fast and efficient mass transit systems have been developed to accommodate the growing long-distance commuting demands between suburbs and central urban areas, as well as between neighboring cities. These transit lines often feature stations spaced several kilometers apart. One example is the Bay Area Rapid Transit (BART) system, which connects dozens of small cities in the East Bay and San Francisco Peninsula to the downtown areas of San Francisco and Oakland, with station spacings up to 8 km (https://www.bart.gov/; https://www.google.com/maps). Other examples include the rising number of satellite cities supported by China's booming high-speed rail network (e.g., the city of Langfang, situated 60 km from Beijing and only a 20-minute ride by high-speed train). A station on these commuter rail lines serves a vast catchment area where most patrons live beyond walking distance from the rail station, and the demand is unevenly distributed. Consequently, designing efficient feeder service networks is crucial for accommodating patrons' first- and last-mile trips, ensuring seamless connectivity and enhancing the overall effectiveness of mass transit systems.

This paper focuses on the optimal design of fixed-route feeder services under spatially heterogeneous demand. Compared to emerging feeder modes, such as ride sourcing, flex-route transit, and shared bikes, fixed-route feeders offer greater reliability thanks to their fixed service schedules and routes, as well as the benefit of remaining impervious to inclement weather conditions. Furthermore, fixed-route feeders are more affordable and energy-efficient, particularly in operating scenarios with relatively high demand densities (e.g., Chang and Schonfeld, 1991).

Two types of models have been commonly employed to address the feeder service optimization problem: discrete models and continuous models. Discrete models incorporate numerous details representing specific physical and geographical constraints in the operating environment, such as the exact locations of candidate stops (e.g., Martins and Pato, 1998; Shrivastava and O'Mahony, 2006; Ciaffi et al., 2012; Lee et al., 2021). These models are often solved using heuristic methods (e.g., Fan and Ran, 2021), making it challenging to gauge the quality of the generated solutions. In contrast, continuous models are more parsimonious, utilizing only a few parameters and design variables to represent the feeder system. Specifically, continuous approximation (CA) models approximate heterogeneous discrete details with continuous functions for demand density, line and stop spacings, and so on. This makes them particularly well-suited for developing optimal transit network layouts under heterogeneous demand (Ouyang et al., 2014; Chen et al., 2015; Chen et al., 2018; Luo et al., 2021). Continuous models are often solved to global optimality or near optimality using efficient analytical or numerical methods. Notable studies in this field are summarized in Table 1 and compared with our research.

Most works in this field assumed spatially uniform demands (Chang and Schonfeld, 1991; Kim and Schonfeld, 2012, 2013, 2014; Sivakumaran et al., 2014; Kim and Schonfeld, 2015, Guo et al., 2018; Su and Fan, 2019; Badia and Jenelius, 2020, 2021). Thus, they cannot be directly applied to real-world feeder system designs under spatially heterogeneous demand. Spatial heterogeneity in demand cannot be overlooked when the feeder's service region is large, as exemplified by the satellite cities mentioned earlier. Most of the cited works also assumed uniform line spacing and headway, while largely neglecting stop spacing optimization. Moreover, they typically assumed fixed dwell time and transfer time between trunk and feeder services, disregarding the relationship between dwell and transfer times and the number of boarding and alighting patrons. Notably, Chang and Schonfeld (1991) and Kim and Schonfeld (2012, 2013, 2014, 2015) considered vehicle size optimization.

Only a few studies have developed feeder network design models for the more realistic, spatially heterogeneous demand. Among them, Wirasinghe (1980) and Sivakumaran et al. (2012) assumed demand is heterogeneous in only one direction (either along the rail line or perpendicular to it), while Kuah and Perl (1988) and Yang et al. (2020) considered demand heterogeneity in both directions of the service region. Most of these studies did not optimize stop spacings, likely due to mathematical complexity. For instance, Quadrifoglio and Li (2009) stated that deriving the optimal stop spacing was "often quite hard." To the best of our knowledge, Kuah and Perl (1988) is the only study that optimized stop spacings under heterogeneous demand. However, their model relied on oversimplified and somewhat unrealistic assumptions (probably necessary to make the problem mathematically tractable), such as very small rail station spacing and ignoring the operating cost associated with bus dwell times. In addition, nearly all these works assumed fixed dwell and transfer times and unlimited feeder bus capacity.

Table 1. Selected research works on fixed-route feeder network design using continuous models

| Research works | Spatial demand pattern | Decision variables | Dwell and transfer times | Capacity constraint | Variable vehicle size | Schedule coordination |
|---|---|---|---|---|---|---|
| Chang and Schonfeld (1991); Kim and Schonfeld (2012, 2013, 2015); Guo et al. (2017); Badia and Jenelius (2020, 2021) | Uniform | Uniform headway and line spacing | Fixed | Yes | Yes | No |
| Kim and Schonfeld (2014) | Uniform | Uniform headway and line spacing | Fixed | Yes | Yes | Yes |
| Quadrifoglio and Li (2009); Li and Quadrifoglio (2010) | Uniform | – | Fixed | No | No | No |
| Sivakumaran et al. (2014) | Uniform | Uniform headway and line spacing | Fixed | Yes | No | No |
| Su and Fan (2019) | Uniform | Spatially-heterogeneous headways, line and stop spacings | Dwell time is a function of boarding number; transfer time is fixed | Yes | No | No |
| Wirasinghe (1980) | Heterogeneous in one direction only | Spatially-heterogeneous headways and line spacings | Fixed | No | No | No |
| Sivakumaran et al. (2012) | Heterogeneous in one direction only | Uniform headway and line spacing | Fixed | No | No | Yes |
| Kuah and Perl (1988) | Heterogeneous | Spatially-heterogeneous headways, line and stop spacings | Fixed | No | No | No |
| Yang et al. (2020) | Heterogeneous | Spatially-heterogeneous headways and line spacings | Fixed | Yes | No | Yes |
| **Our work** | **Heterogeneous** | **Spatially-heterogeneous headways, line and stop spacings** | **Functions of boarding and alighting numbers** | **Yes** | **Yes** | **Yes** |

To summarize, despite the extensive history of literature on fixed-route feeder service design, the following research gaps have not yet been adequately addressed:

(i) The optimization of heterogeneous stop spacings has been largely neglected in the literature. Most studies assume that the stop spacing is a constant or do not explicitly model stop locations and bus dwell times. Kuah and Perl (1988) is the only work that optimized heterogeneous stop spacings under heterogeneous demand, but their model also has its drawbacks, as discussed earlier in this paper. Note that real-world stop spacings vary significantly; for example, Texas Transportation Institute (1996) states that real-world stop spacings range from 200 m to 800 m. And they substantially affect passengers' access distances and in-vehicle travel times (Daganzo and Ouyang, 2019). Therefore, stop spacings should be jointly optimized with line spacings and headways in a heterogeneous manner.

(ii) Most studies assume constant dwell times and transfer times for feeder bus services, oversimplifying the impact of passenger boarding and alighting on dwell times. Su and Fan (2019) is the only exception, but it assumes uniform demand. Unlike rail transit, bus dwell time mainly depends on the number of passengers getting on and off at a stop (Daganzo and Ouyang, 2019). Ignoring this relationship can render errors in estimating bus roundtrip times and in-vehicle travel times for passengers. Particularly, when bus size is jointly optimized, employing a fleet of smaller vehicles can result in lower operating costs, increased service frequencies, and reduced travel times and roundtrip durations due to the fewer passengers each vehicle carries. This trade-off and its impact on the overall system cost cannot be accurately represented if the relationship between the number of boarding and alighting passengers and the dwell times is not taken into account.

(iii) While previous studies (e.g., Sivakumaran et al., 2012; Kim and Schonfeld, 2014; Yang et al., 2020) have investigated the advantages of coordinating schedules between trunk and feeder services, their conclusions are hindered by modeling limitations. These works, for instance, overlooked stop spacing and variable dwell times, assumed uniform line spacing and headway, or generated suboptimal solutions. In addition, these studies only focused on coordination in one direction (the patron collection direction), while the coordination effects in the patron distribution direction were ignored. As we will see soon, schedule coordination in the patron distribution direction provides a smaller yet noteworthy benefit. The overall benefit of schedule coordination is significantly greater than previously predicted in the literature.

Besides, a simple but important question seems to be neglected in literature. Specifically, for a rectangular or approximately rectangular feeder service region (which is often found in grid trunk-line networks; see Chien and Schonfeld, 1997), which feeder line layout is more cost-effective: each line making stops along the shorter side of the rectangle and traveling nonstop along the longer side, or vice versa? In this paper, we will demonstrate that the difference between the two layout options is sizeable.

In summary, existing models lack the comprehensive nature required for accurately designing a fixed-route feeder system and evaluating its cost-effectiveness. More inclusive models are essential for assisting decision-makers to select the most suitable feeder network design for a trunk-line rail system. Firstly, the models should reveal the full cost and benefit of fixed-route feeders, providing a sound basis for comparing against alternative feeder modes, such as flex-route feeder, ride-hailing, and shared bikes. Secondly, they need to offer detailed guidance on the optimal design of a fixed-route feeder network, including line layout, stop locations, and service headways. To this end, we formulate more comprehensive CA models to explore the optimal fixed-route feeder design problem under spatially heterogeneous demand. The models aim to minimize the sum of the feeder system's operating cost and the feeder patrons' travel cost, as well as their transfer cost at the rail terminal. They account for the impacts of boarding and alighting passengers on dwell times and transfer times, and optimize the spatially varied stop spacings for each feeder line[1], as well as the line spacings, headways, and vehicle size. Additionally, our models can effectively harness the greater potential of schedule coordination by incorporating coordination efforts in both patron collection and distribution directions.

The models are solved by combining the first-order condition results with an iterative search algorithm. Extensive numerical analyses are performed to examine: (i) the accuracy of the CA models; (ii) the properties of the optimal design; (iii) the sensitivity of cost savings to the value of time, demand pattern, demand rate, service region size, and alternative feeder line layouts; and (iv) the effects of schedule coordination for each service direction individually, as well as for both directions simultaneously. The results unveil new insights with practical implications.

The main contributions of this paper are summarized as follows:

(i) To our best knowledge, we are the first to model the optimal fixed-route feeder network design with heterogeneous stop spacings under heterogeneous demand by incorporating more realistic dwell time and transfer delay models and the vehicle capacity constraint. Our results indicate that the system cost reduction by considering heterogeneous stop spacings is moderate but consistent, and is larger when the demand is more heterogeneous.

(ii) We explore the fuller benefit of schedule coordination between trunk and feeder services in both service directions, acknowledging the asymmetry in schedule coordination between the two service directions.

(iii) Through numerical analyses, we demonstrate how various operating factors (e.g., the demand's spatial pattern, demand rate, and service region size) affect the optimal feeder network design and the cost savings compared to homogeneous designs and designs not considering heterogeneous stop spacings.

In addition, we compare two alternative feeder line layouts and find that it consistently results in lower costs for feeder buses to travel along the shorter side of the service region to make stops and along the longer side to the rail station without stopping.

The rest of the paper is structured as follows. Section 2 presents the problem setup and the CA model formulation. Analytical properties of the optimal solution and the solution algorithm are furnished in Section 3. Section 4 discusses the numerical case study results and findings. Finally, conclusions and future work are outlined in Section 5.

## 2. Methodology

Assumptions and settings of our feeder design problem are presented in Section 2.1. The cost models are formulated in Sections 2.2 and 2.3. Section 2.4 furnishes the optimization model without schedule coordination. Schedule coordination between the trunk line service and the feeders is modeled in Section 2.5.

---

[1] Incorporating the optimization of heterogeneous stop spacing increases the modeling complexity, as our CA models involve a bivariate decision function (i.e., stop spacing is a function of two spatial coordinates, $x$ and $y$) and double integrals of that decision function; see Section 2.

## 2.1 Problem setup

Figure 1 depicts a schematic representation of a feeder network. A feeder bus terminal, marked by the black dot, is located at the left-bottom corner of the figure. This terminal can be a trunk-line (e.g., rail transit) station in a city's suburban area or an intercity rail station.[2] It is assumed that the terminal's catchment area, which represents the region served by the feeder network, is a rectangular zone centered at the terminal, encompassing a dense street grid. A *quarter* of this catchment area is defined as $\Omega \equiv \{(x,y)|0 \leq x \leq L, 0 \leq y \leq W\}$; see Fig. 1.

Owing to the symmetric nature of the area, it is only necessary to identify the optimal design for feeder lines carrying patrons within $\Omega$, both to and from the terminal. We presume that feeder buses collect or distribute passengers along the $y$-axis, while they undertake nonstop travel along the $x$-axis either towards or away from the terminal, as illustrated by the thin solid lines in the figure. The positions of feeder bus stops are indicated by black squares.

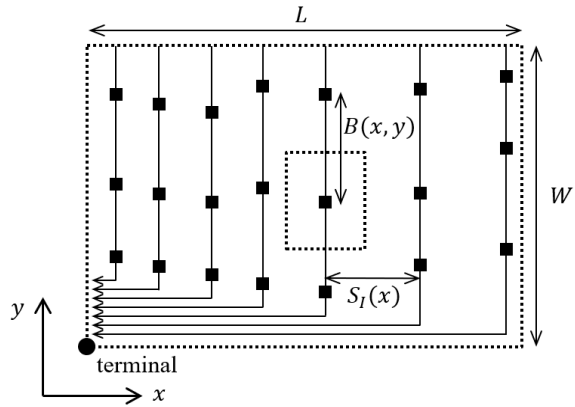

**Fig. 1 The feeder system layout**

Travel demands in this service area are expressed as continuous, *time-invariant* density functions.[3] Specifically, the demand density from location $(x,y) \in \Omega$ to the terminal (in the patron-collection direction) is denoted by $\lambda_p(x,y)$, while the demand density from the terminal to $(x,y) \in \Omega$ (in the patron-distribution direction) is denoted by $\lambda_d(x,y)$. The line spacing, $S_I(x)$, and the headways in the patron-collection and distribution directions, $H_{Ip}(x)$ and $H_{Id}(x)$ respectively, are functions of $x$. The stop spacing, $B(x,y)$, is a function of both $x$ and $y$. In the context of the CA framework, these decision functions are assumed to be integrable on $0 \leq x \leq L$ and $\Omega$, respectively.

We further assume that: (i) each patron chooses the nearest feeder stop on the closest line for boarding or alighting; (ii) feeder buses on any specific line maintain regular headways; (iii) patrons arrive at the feeder stops randomly without consulting the service timetable; and (iv) all patrons waiting at the terminal or a feeder bus stop will be picked up by the next arriving bus (i.e., no patron will be left behind). These assumptions are commonly employed in the literature (e.g., Daganzo, 2010; Chen et al., 2015; Fan et al., 2018).

---

[2] The terminal may also represent a central business district in a medium or small-sized city, a shopping center within an urban district, or an airport. In such instances, the feeder service effectively functions as a shuttle service. Nevertheless, our models remain applicable to these cases, requiring only minor modifications.

[3] The models can readily be adapted to real-world situations where demand densities fluctuate over time, e.g., between peak and off-peak periods. To account for these variations, one can model the costs associated with different time periods and aggregate them accordingly; see Wu et al. (2020) for an example.

The problem aims to minimize the *generalized cost* of the feeder system, consisting of the patrons' travel cost and the agency's operating cost. All cost components are expressed in units of hours; in other words, the agency cost items originally expressed in monetary units are converted to hours to ensure they can be added together with the patron costs. The following two sections develop these cost terms incurred during one hour of bus operations.

**2.2 Patrons' travel time cost**

The patrons' travel time cost comprises three components: (i) the time spent accessing (for the patron-collection direction) and egressing (for the distribution direction) feeder stops; (ii) the waiting time at feeder stops and the transfer time at the terminal; and (iii) the in-vehicle travel time on feeder buses. They are formulated in Sections 2.2.1 to 2.2.3, respectively.[4] Note that while bus fare is another component of patron cost, it is excluded from our generalized cost model. This exclusion is due to the fare constituting a monetary transfer from patrons to the transit agency, thereby representing a cost for patrons and a revenue source for the agency.

*2.2.1 Access and egress cost*

The total access and egress time per hour of bus operations, denoted by $C_A$, is formulated as follows:

$$C_A = \int_{x=0}^{L} \int_{y=0}^{W} \frac{S_I(x) + B(x,y)}{4v_W} [\lambda_p(x,y) + \lambda_d(x,y)] \, dy \, dx \tag{1}$$

where $(S_I(x) + B(x,y))/4$ represents the average access or egress distance per passenger originating from or destined for location $(x,y)$; and $v_W$ denotes the walking speed.[5] Note that the double integral part related to $B(x,y)$ would not appear (i.e., the RHS of (1) would reduce to a single integral of decision function $S_I(x)$) if stop spacings are not jointly optimized.

*2.2.2 Waiting and transfer cost*

The total waiting and transfer time per hour of bus operations, denoted by $C_W$, is formulated as:

$$C_W = C_{Wp} + C_{Wd} \tag{2}$$

---

[4] In this paper, the total patron cost is formulated by summing the three patron travel time components. Alternatively, different weighting coefficients could be applied to these components to account for the varying values of time during different stages of travel.

[5] Eq. (1) is based upon the assumption that a patron always chooses the nearest feeder line and the nearest feeder stop on that line for boarding or alighting (see assumption (i) in Section 2.1). In a heterogeneous feeder network, a patron's nearest stop might not be situated on the closest line. Nevertheless, incorporating the choice of the nearest stop would considerably complicate the modeling process, while only moderately improving model accuracy. Thus, we opt for this simple (and conservative) approach for modeling patrons' stop selection.

$$C_{Wp} = \int_{x=0}^{L}\int_{y=0}^{W} \frac{H_{Ip}(x)}{2}\lambda_p(x,y)dy\,dx + \int_{x=0}^{L}\int_{y=0}^{W}\left(\frac{\tau_a}{2}S_I(x)H_{Ip}(x)\cdot\int_{y=0}^{W}\lambda_p(x,y)dy + t_{f-t} + \frac{H_t}{2}\right)\lambda_p(x,y)dy\,dx \quad (3)$$

$$C_{Wd} = \int_{x=0}^{L}\int_{y=0}^{W}\left(t_{t-f} + \frac{H_{Id}(x)}{2} + \frac{\tau_b}{2}S_I(x)H_t\cdot\int_{y=0}^{W}\lambda_d(x,y)dy\right)\lambda_d(x,y)dy\,dx \quad (4)$$

where $C_{Wp}$ and $C_{Wd}$ denote the total waiting and transfer times for patrons traveling to and from the terminal, respectively. The first term on the RHS of (3) corresponds to the total waiting time at the feeder stops, with $H_{Ip}(x)/2$ being the average waiting time per patron for the bus line located at $x$. The second term on the RHS of (3) accounts for the total patron delay at the terminal, which comprises three components: the time lost due to patrons' alighting process from feeder buses, the transfer delay from feeder to trunk-line transit, and the waiting time for a trunk-line vehicle. Here, $\tau_a$ denotes the alighting time per patron, $t_{f-t}$ the transfer delay per patron (including walking time), and $H_t$ the trunk-line service headway at the terminal. These components are illustrated in Fig. 2. Note in particular that $S_I(x)H_{Ip}(x)\int_{y=0}^{W}\lambda_p(x,y)dy$ represents the number of onboard patrons per bus on a line located at $x$, while $\frac{\tau_a}{2}S_I(x)H_{Ip}(x)\int_{y=0}^{W}\lambda_p(x,y)dy$ corresponds to the average time loss per patron during the alighting process at the terminal. Similarly, the three components on the RHS of (4) are: the transfer delay from trunk-line to feeder, the waiting time for the feeder service, and the boarding time loss, where $t_{t-f}$ denotes the transfer delay per patron and $\tau_b$ the boarding time per patron.

The boarding and alighting time losses in (3) and (4) are significant, as all patrons will board or alight at the terminal, resulting in the longest bus dwell time at this location. Regrettably, these factors were overlooked in previous feeder design models (e.g., Chang et al., 1991; Sivakumaran et al., 2014; Su et al., 2019; Badia et al., 2020; Yang et al., 2020).

Additionally, note that the formulas for $C_{Wp}$ and $C_{Wd}$ are asymmetrical. Thus, modeling a single direction of travel or two symmetrical directions (e.g., Chang and Schonfeld, 1991; Kim and Schonfeld, 2012; Sivakumaran et al., 2012) is insufficient for optimal feeder system design.

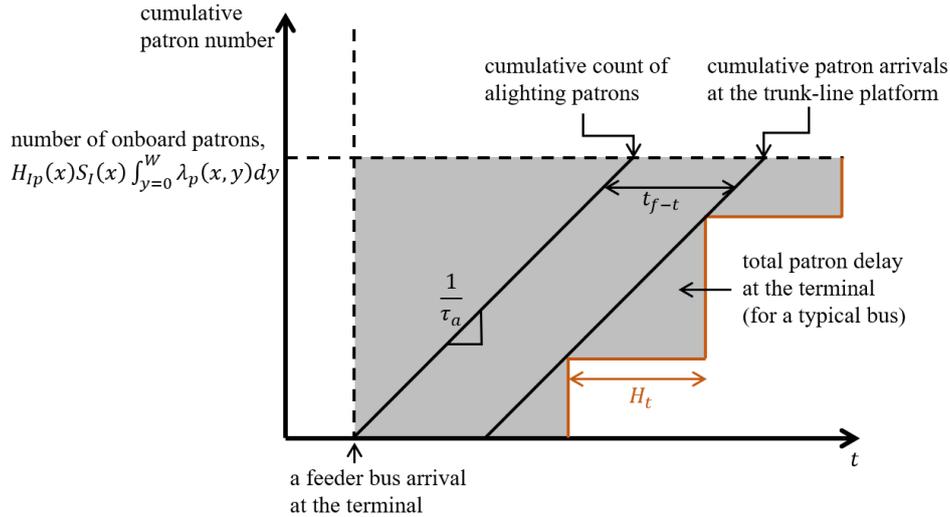

**Fig. 2 Patron delays at the terminal (in the patron-collection direction)**

### 2.2.3 In-vehicle travel cost

The total in-vehicle travel time per operation hour, $C_T$, is composed of three parts: (i) the time to overcome the distance, $C_{T1}$; (ii) the time lost to bus deceleration, acceleration, door opening and closing at stops, $C_{T2}$; and (iii) the time lost to patrons' boarding and alighting behavior, $C_{T3}$. Formulations of these cost items are given in (5)–(8).

$$C_T = C_{T1} + C_{T2} + C_{T3} \tag{5}$$

$$C_{T1} = \int_{x=0}^{L} \int_{y=0}^{W} \frac{(x+y)}{v_I} [\lambda_p(x,y) + \lambda_d(x,y)] dy dx \tag{6}$$

$$C_{T2} = \tau_0 \int_{x=0}^{L} \int_{y=0}^{W} \frac{\int_{z=y}^{W} (\lambda_p(x,z) + \lambda_d(x,z)) dz}{B(x,y)} dy dx \tag{7}$$

$$C_{T3} = \int_{x=0}^{L} \int_{y=0}^{W} \Big( \tau_b S_I(x) H_{Ip}(x) \lambda_p(x,y) \int_{z=y}^{W} \lambda_p(x,z) dz +$$

$$\tau_a S_I(x) H_{Id}(x) \lambda_d(x,y) \int_{z=y}^{W} \lambda_d(x,z) dz \Big) dy dx \tag{8}$$

where $v_I$ denotes the cruise speed of feeder buses; and $\tau_0$ the fixed time loss per bus at a stop due to bus deceleration, acceleration, door opening and closing. In (7), $\int_{z=y}^{W} \lambda_p(x,z) dz \cdot dx$ and $\int_{z=y}^{W} \lambda_d(x,z) dz \cdot dx$ represent the onboard patron flows passing $y$ within a swath of width $dx$ for the collection and distribution directions, respectively.

The first term on the RHS of (8) is derived as follows. First, $S_I(x) H_{Ip}(x) \int_{z=y}^{W} \lambda_p(x,z) dz$ represents the number of patrons onboard a bus on the line located at $x$ when the bus passes $y$ in the collection direction. Second, $\tau_b \lambda_p(x,y) dy dx$ corresponds to the boarding time loss generated by the patrons originating from an area of size $dx \times dy$ at $(x,y)$. This time loss is then multiplied by $S_I(x) H_{Ip}(x) \int_{z=y}^{W} \lambda_p(x,z) dz$ since it affects all onboard patrons of a bus passing $(x,y)$. The second term on the RHS of (8) is developed in a similar manner. Note again that (7) would not involve a double integral of decision function $B(x,y)$ if stop spacings were not jointly optimized. Moreover, (8) would not exist if the dependency of bus dwell times on the boarding and alighting patrons were disregarded.

## 2.3 Agency cost

During each hour of operation, the feeder service operator incurs three types of costs: (i) the stop infrastructure cost, denoted by $C_s$; (ii) the operating cost associated with bus-km traveled (e.g., fuel cost), $C_{vk}$; and (iii) the cost related to the bus fleet, including amortized bus purchase and maintenance costs, as well as driver wages[6], $C_{vh}$. Note that these costs are converted to hours using the average value of time for patrons, $\theta$. The formulations for these costs are as follows:

$$C_s = \frac{\pi_s}{\theta} \int_{x=0}^{L} \int_{y=0}^{W} \frac{1}{S_I(x) B(x,y)} dx dy \tag{9}$$

$$C_{vk} = \frac{\pi_v}{\theta} \int_{x=0}^{L} \left[ \frac{(W+x)}{S_I(x)} \left( \frac{1}{H_{Ip}(x)} + \frac{1}{H_{Id}(x)} \right) \right] dx \tag{10}$$

$$C_{vh} = \frac{\pi_m}{\theta} V_h \tag{11}$$

where $\pi_s$, $\pi_v$, and $\pi_m$ denote the unit costs per feeder stop ($/stop), bus-km traveled ($/bus-km), and bus hour ($/bus-hour), respectively. Note that $\frac{1}{S_I(x) B(x,y)} dx dy$ is the number of stops in the area of

---

[6] In some studies (e.g., Gu et al., 2016), this cost component is alternatively referred to as the time-based or bus-hour based operating cost.

size $dx \times dy$ at $(x, y)$, and $\frac{(W+x)}{S_I(x)}dx$ is the bus line length contributed by a "vertical" swath of width $dx$ at $x$. The $V_h$ in (11) denotes the required bus fleet size, which comprises (i) the number of buses that are actively travelling between stops, $V_{h1}$; and (ii) the number of buses that are dwelling at stops. The latter component can be further divided into the number of buses needed due to deceleration, acceleration, door opening and closing at stops, $V_{h2}$, as well as the number of buses required for passenger boarding and alighting, $V_{h3}$. The three components $V_{h1}$, $V_{h2}$, and $V_{h3}$ are formulated in (12)–(14), respectively:

$$V_{h1} = \int_{x=0}^{L} \left[ \frac{(W+x)}{S_I(x)v_I} \left( \frac{1}{H_{Ip}(x)} + \frac{1}{H_{Id}(x)} \right) \right] dx \tag{12}$$

$$V_{h2} = \int_{x=0}^{L} \int_{y=0}^{W} \frac{\tau_0}{S_I(x)B(x,y)} \left( \frac{1}{H_{Ip}(x)} + \frac{1}{H_{Id}(x)} \right) dxdy \tag{13}$$

$$V_{h3} = (\tau_a + \tau_b) \int_{x=0}^{L} \int_{y=0}^{W} [\lambda_p(x,y) + \lambda_d(x,y)] dydx \tag{14}$$

Derivation of (12)–(14) is similar to those presented earlier in this section.

## 2.4 Optimization model

The generalized cost can be expressed as the sum of all the above cost components:

$$GC = C_S + C_{vk} + C_{vh} + C_A + C_W + C_T \tag{15a}$$

The optimization problem is formulated as follows:
$$\min GC \tag{15b}$$
subject to:
$$\int_{y=0}^{W} \lambda_p(x,y) dy \cdot S_I(x) H_{Ip}(x) \leq K \tag{15c}$$
$$\int_{y=0}^{W} \lambda_d(x,y) dy \cdot S_I(x) H_{Id}(x) \leq K \tag{15d}$$
$$H_{min} \leq H_{Ip}(x) \leq H_{max} \tag{15e}$$
$$\max\{H_{min}, H_t\} \leq H_{Id}(x) \leq H_{max} \tag{15f}$$

where $K$ denotes a feeder bus's patron-carrying capacity; and $H_{min}$, $H_{max}$ the minimum and maximum headways, respectively. Constraints (15c) and (15d) ensure that the feeder bus capacity is sufficient for carrying the patrons. Constraints (15e)–(15f) are boundary constraints. Note that the feeder headway in the distribution direction, $H_{Id}(x)$, must be greater than or equal to the trunk-line headway, $H_t$. Otherwise, some feeder buses would not have any passengers to collect at the terminal.

## 2.5 Modeling schedule coordination

Feeder and trunk-line schedules can be coordinated to minimize waiting times during transfers (Sivakumaran et al., 2012). Coordination in both patron-collection and distribution directions can effectively eliminate the waiting time components, $\frac{H_t}{2}$ in Eq. (3) and $\frac{H_{Id}(x)}{2}$ in Eq. (4). However, the coordination schemes in the two directions are different. In the patron-collection direction, coordination necessitates that a trunk-line vehicle is dwelling at the terminal when patrons alighting from a feeder bus arrive at the trunk platform. This is only feasible if trunk-line vehicles bound for all destinations arrive at the terminal simultaneously. This may occur when a single trunk line serves the terminal and most transferring patrons travel in the same direction (e.g., during rush hour commuting). Meanwhile, coordination in the distribution direction merely requires a feeder bus to be ready for departure when patrons disembarking a trunk-line vehicle arrive at the feeder stop. In this case, coordination can be accomplished even when the terminal serves multiple trunk lines operating on distinct schedules.

For simplicity, we assume that the terminal serves only one trunk line, with patrons traveling to and from a city center using this line. This assumption facilitates coordination between feeder and trunk-line services. Specifically, to achieve coordination in the collection direction, a feeder line's headway must be an integer multiple of $H_t$, i.e.,

$$H_{Ip}(x) = kH_t \ (k = 1, 2, 3, ...); \tag{16}$$

while in the distribution direction, all feeder-line headways must equal $H_t$, i.e.,

$$H_{Id}(x) = H_t, \ \forall x \in [0, L]. \tag{17}$$

This highlights the asymmetry in the two travel directions when coordination is implemented. We further assume that feeder buses always arrive on schedule and that feeder bus layover time costs are disregarded (e.g., Kim and Schonfeld, 2014; Mei et al., 2021).

Under schedule coordination, the total waiting and transfer times in both travel directions, originally represented by Eqs. (3) and (4), are replaced by the following equations:

$$C_{Wp}' = \int_{x=0}^{L} \int_{y=0}^{W} \frac{H_{Ip}(x)}{2} \lambda_p(x,y) dy\, dx + \int_{x=0}^{L} \int_{y=0}^{W} \left( \tau_a S_I(x) H_{Ip}(x) \cdot \int_{y=0}^{W} \lambda_p(x,y) dy + t_{f-t} \right) \lambda_p(x,y) dy\, dx \tag{18}$$

$$C_{Wd}' = \int_{x=0}^{L} \int_{y=0}^{W} \left( t_{t-f} + \tau_b S_I(x) H_t \cdot \int_{y=0}^{W} \lambda_d(x,y) dy \right) \lambda_d(x,y) dy\, dx \tag{19}$$

In Eq. (18), the second term consists of the total patron alighting time loss and the total transfer delay. Note that the waiting time for the trunk-line vehicle, $\frac{H_t}{2}$, is eliminated due to the coordination, and the alighting time loss term in Eq. (3) is doubled. The derivation is illustrated in Fig. 3 for a case of $k = 1$. Eq. (19) is developed in a similar manner.

The remaining part of the optimization model involving schedule coordination is identical to (15a)–(15f).

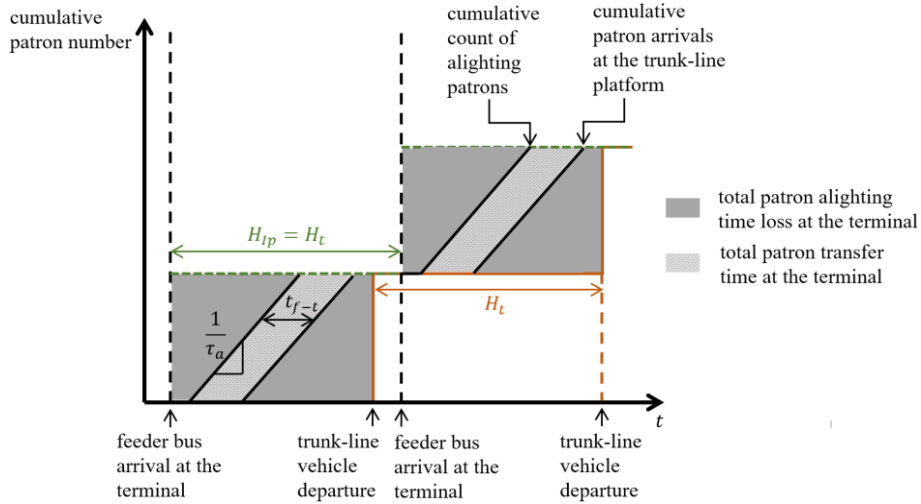

Fig. 3 Patron delays at the terminal under schedule coordination with $k = 1$ (the collection direction)

## 3. Solution method

For brevity, this section only focuses on the solution approach for (15a)–(15f). The model involving schedule coordination (i.e., Eq. (18) and (19)) can be solved in a highly similar fashion. Section 3.1 derives the properties characterizing a globally optimal solution. Section 3.2 furnishes a heuristic solution algorithm built upon these optimality properties.

### 3.1 Optimality properties

To simplify the presentation, we begin by defining the following aggregate demand quantities and functions:

$$\Lambda = \Lambda_p + \Lambda_d; \tag{20a}$$
$$\Lambda_p = \int_{x=0}^{L} \int_{y=0}^{W} \lambda_p(x,y) dy dx; \tag{20b}$$
$$\Lambda_d = \int_{x=0}^{L} \int_{y=0}^{W} \lambda_d(x,y) dy dx; \tag{20c}$$
$$\Lambda_{px}(x) = \int_{y=0}^{W} \lambda_p(x,y) \, dy, \ 0 \le x \le L; \tag{20d}$$
$$\Lambda_{dx}(x) = \int_{y=0}^{W} \lambda_d(x,y) \, dy, \ 0 \le x \le L; \tag{20e}$$
$$\Lambda_{pxy}(x,y) = \int_{z=y}^{W} \lambda_p(x,z) dz, 0 \le x \le L, 0 \le y \le W; \tag{20f}$$
$$\Lambda_{dxy}(x,y) = \int_{z=y}^{W} \lambda_d(x,z) dz, 0 \le x \le L, 0 \le y \le W; \tag{20g}$$
$$M_{px}(x) = \int_{y=0}^{W} \lambda_p(x,y) \Lambda_{pxy}(x,y) dy, 0 \le x \le L; \tag{20h}$$
$$M_{dx}(x) = \int_{y=0}^{W} \lambda_d(x,y) \Lambda_{dxy}(x,y) dy, 0 \le x \le L. \tag{20i}$$

where $\Lambda$, $\Lambda_p$, and $\Lambda_d$ represent the total hourly demand for both directions combined, the collection direction, and the distribution direction, respectively; $\Lambda_{px}(x)$ and $\Lambda_{dx}(x)$ the demand densities integrated over the $y$-direction at location $x$ for collection and distribution travels, respectively; and $\Lambda_{pxy}(x,y)$ and $\Lambda_{dxy}(x,y)$ the demand densities at location $x$, integrated from $y$ to the top edge of $\Omega$, for collection and distribution travels, respectively (note that $\Lambda_{pxy}(x,0) = \Lambda_{px}(x)$ and $\Lambda_{dxy}(x,0) = \Lambda_{dx}(x)$). The $\Lambda_{pxy}(x,y)$ and $\Lambda_{dxy}(x,y)$ are utilized to calculate the onboard patron numbers at $y$ for a bus on a line located at $x$. Lastly, $M_{px}(x)$ and $M_{dx}(x)$ are employed to calculate the total dwell time loss for all patrons combined.

The following functions are defined for further simplification of the optimality properties:

$$\alpha(x) = \frac{1}{\theta}\left(\pi_v(W+x) + \frac{\pi_m(W+x)}{v_I} + \pi_m \tau_0 \int_{y=0}^{W} \frac{1}{B(x,y)} dy\right), 0 \le x \le L; \tag{21a}$$
$$\beta_p(x) = \tau_a[\Lambda_{px}(x)]^2 + 2\tau_b M_{px}(x), 0 \le x \le L; \tag{21b}$$
$$\beta_d(x) = 2\tau_a M_{dx}(x), 0 \le x \le L; \tag{21c}$$
$$\gamma(x) = \frac{1}{\theta}\pi_s \int_{y=0}^{W} \frac{1}{B(x,y)} dy, 0 \le x \le L. \tag{21d}$$

Here $\alpha(x)$ denotes the agency cost rate per bus on a line located at $x$, excluding the agency cost associated with boarding and alighting patrons; $\beta_p(x)$ and $\beta_d(x)$ are related to the boarding and alighting time loss in the collection and distribution travel directions, respectively; and $\gamma(x)$ represents the stop infrastructure cost rate for a line located at $x$.

Note that $\Lambda_{pxy}(x,y)$, $\Lambda_{dxy}(x,y)$, $M_{px}(x)$, $M_{dx}(x)$, $\beta_p(x)$, and $\beta_d(x)$ are needed because the impacts of boarding and alighting patrons on bus dwell times and transfer delays are explicitly modeled.

The optimal solution to (15a)–(15f) must satisfy the following analytical properties for any $x \in [0, L]$ and $y \in [0, W]$. The derivation of these properties can be found in Appendix B.

$$H_{Ip}^*(x) = mid \left\{ H_{min}, \min\left\{\frac{K}{\Lambda_{px}(x)S_I^*(x)}, H_{max}\right\}, \sqrt{\frac{2\alpha(x)\left(S_I^*(x)\right)^{-1}}{\Lambda_{px}(x)+\beta_p(x)S_I^*(x)}} \right\} \tag{22a}$$

$$H_{Id}^*(x) = mid \left\{ \max\{H_{min}, H_t\}, \min\left\{\frac{K}{\Lambda_{dx}(x)S_I^*(x)}, H_{max}\right\}, \sqrt{\frac{2\alpha(x)\left(S_I^*(x)\right)^{-1}}{\Lambda_{dx}(x)+\beta_d(x)S_I^*(x)}} \right\} \tag{22b}$$

$$S_I^*(x) = \min\left\{ \sqrt{\frac{2\alpha(x)\left(\left(H_{Ip}^*(x)\right)^{-1}+\left(H_{Id}^*(x)\right)^{-1}\right)+2\gamma(x)}{\frac{\Lambda_{px}(x)+\Lambda_{dx}(x)}{2v_W}+\tau_b[\Lambda_{dx}(x)]^2 H_t+\beta_p(x)H_{Ip}^*(x)+\beta_d(x)H_{Id}^*(x)}}, \frac{K}{\Lambda_{px}(x)H_{Ip}^*(x)}, \frac{K}{\Lambda_{dx}(x)H_{Id}^*(x)} \right\} \tag{22c}$$

$$B^*(x,y) = \sqrt{\frac{4v_w\left[\frac{1}{\mu S_I^*(x)}\left(\pi_s+\frac{\pi_m\tau_0}{H_{Ip}^*(x)}+\frac{\pi_m\tau_0}{H_{Id}^*(x)}\right)+\tau_0\left(\Lambda_{pxy}(x,y)+\Lambda_{dxy}(x,y)\right)\right]}{\lambda_p(x,y)+\lambda_d(x,y)}} \tag{22d}$$

Eqs. (22a)–(22c) show that the optimal headways and line spacings are negatively correlated with demand and positively correlated with the agency cost rate, $\alpha(x)$. In other words, lower demand (or higher agency cost rates) leads to larger bus service headways and line spacings. Note if the relationship between boarding and alighting patrons and bus dwell times is ignored, $\tau_0$ and $\alpha(x)$ would be larger, resulting in an overestimation of optimal headways and line spacings.

Eq. (22d) reveals that the optimal stop spacing is positively correlated with the agency cost rates and the number of onboard patrons (represented by $\Lambda_{pxy}(x,y) + \Lambda_{dxy}(x,y)$). But it is negatively correlated with the local demand density $\lambda_p(x,y) + \lambda_d(x,y)$ since shorter stop spacings can reduce access and egress costs. Notably, the optimal heterogeneous stop spacing is directly related to the local line spacing and headway, i.e., longer line spacing and headways result in shorter stop spacings. This highlights the importance of jointly and properly optimizing heterogeneous stop spacing. Eq. (22d) also shows that stop spacing would be overestimated if the impacts of boarding and alighting patrons on bus dwell times are ignored, thanks to the larger $\tau_0$ that would occur.

Based on the above properties, we propose the following iterative algorithm to heuristically solve the optimization problem.

### 3.2 A heuristic iterative solution method

The algorithm consists of two main stages (i.e., Stage 1 and 2 in the following algorithm) that iterate until convergence is reached. In the first stage, we develop $B^*(x,y)$ using (22d) and given values of $H_{Ip}(x)$, $H_{Id}(x)$, and $S_I(x)$ for a lattice of discrete points defined on $\Omega$. In the second stage, $S_I^*(x)$, $H_{Ip}^*(x)$, and $H_{Id}^*(x)$ are derived using (22a)–(22c) given the $B^*(x,y)$ obtained in the first stage.

**Stage 0: Initialization.**

(0.1) Discretize the service region $\Omega$ into a $m \times n$ lattice of points, setting $x_i = (i - 0.5)\frac{L}{n}$, $i = 1,2,\ldots,n$, and $y_j = (j - 0.5)\frac{W}{m}$, $i = 1,2,\ldots,m$.

(0.2) Initialize $H_{Ip}^{(0)}(x_i)$, $H_{Id}^{(0)}(x_i)$, and $S_I^{(0)}(x_i)$ to satisfy the boundary constraints for $i = 1,2,\ldots,n$. Set the outer-loop iteration count $k = 1$ and proceed to Stage 1.

**Stage 1: Compute $B(x, y)$.**

Calculate $B^{(k)}(x_i, y_j)$ using Eq. (22d) for $i = 1,2,\ldots,n$, $j = 1,2,\ldots,m$, taking $H_{Ip}^{(k-1)}(x_i)$, $H_{Id}^{(k-1)}(x_i)$, and $S_I^{(k-1)}(x_i)$ as inputs. Note that the integrals are approximated by summations. Proceed to Stage 2.

**Stage 2: Compute $H_{Ip}(x)$, $H_{Id}(x)$ and $S_I(x)$.**

(2.1) Set $\tilde{S}_I^{(k,0)}(x_i) = S_I^{(k-1)}(x_i)$, $i = 1,2,\ldots,n$. Initialize the inner-loop iteration count $k' = 1$.

(2.2) Taking $B^{(k)}(x_i, y_j)$ and $\tilde{S}_I^{(k,k'-1)}(x_i)$ ($i = 1,2,\ldots,n$, $j = 1,2,\ldots,m$) as inputs, calculate $\tilde{H}_{Ip}^{(k,k')}(x_i)$ and $\tilde{H}_{Id}^{(k,k')}(x_i)$ using Eqs. (22a)–(22b).

(2.3) Taking $B^{(k)}(x_i, y_j)$, $\tilde{H}_{Ip}^{(k,k')}(x_i)$, and $\tilde{H}_{Id}^{(k,k')}(x_i)$ ($i = 1,2,\ldots,n$, $j = 1,2,\ldots,m$) as inputs, calculate $\tilde{S}_I^{(k,k')}(x_i)$ by (22c).

(2.4) Set $S_I^{(k)}(x_i) = \tilde{S}_I^{(k,k')}(x_i)$, $H_{Ip}^{(k)}(x_i) = \tilde{H}_{Ip}^{(k,k')}(x_i)$, and $H_{Id}^{(k)}(x_i) = \tilde{H}_{Id}^{(k,k')}(x_i)$. Check for convergence using the predefined error tolerance $\varepsilon$. If the following convergency criteria are satisfied: $\sum_{i=1}^{n}\left|\tilde{S}_I^{(k,k')}(x_i) - \tilde{S}_I^{(k,k'-1)}(x_i)\right| \leq n\varepsilon$ and $\sum_{i=1}^{n}\left(\left|\tilde{H}_{Ip}^{(k,k')}(x_i) - \tilde{H}_{Ip}^{(k,k'-1)}(x_i)\right| + \left|\tilde{H}_{Id}^{(k,k')}(x_i) - \tilde{H}_{Id}^{(k,k'-1)}(x_i)\right|\right) \leq 2n\varepsilon$, then proceed to the next step. Otherwise, update $k' \leftarrow k' + 1$ and repeat steps (2.2) and (2.3).

(2.5) Check convergency. If $\sum_{i=1}^{n}\left|S_I^{(k)}(x_i) - S_I^{(k-1)}(x_i)\right| \leq n\varepsilon$, $\sum_{i=1}^{n}\left(\left|H_{Ip}^{(k)}(x_i) - H_{Ip}^{(k-1)}(x_i)\right| + \left|H_{Id}^{(k)}(x_i) - H_{Id}^{(k-1)}(x_i)\right|\right) \leq 2n\varepsilon$, and $\sum_{i=1}^{n}\sum_{j=1}^{m}\left|B^{(k)}(x_i, y_j) - B^{(k-1)}(x_i, y_j)\right| \leq mn\varepsilon$ are all satisfied, set $S_I^*(x_i) = S_I^{(k)}(x_i)$, $H_{Ip}^*(x_i) = H_{Ip}^{(k)}(x_i)$, $H_{Id}^*(x_i) = H_{Id}^{(k)}(x_i)$, and $B^*(x_i, y_j) = B^{(k)}(x_i, y_j)$ for $i = 1,2,\ldots,n$, $j = 1,2,\ldots,m$, and end the search. Otherwise, update $k \leftarrow k + 1$ and return to Stage 1.

The optimal CA solution, obtained on the point lattice, can be transformed into a specific layout of feeder lines. The methodology for this conversion is detailed in Appendix C.

## 4. Numerical case studies

Section 4.1 outlines the setup of numerical case studies. The accuracy of our CA models is examined in Section 4.2. Optimal feeder system designs for a typical case in both high- and low-wage cities are presented and discussed in Section 4.3. In Section 4.4, the sensitivity of the optimal feeder design to various key operating factors, such as demand pattern, demand rate, size of the service region, and feeder line layout, is presented. Lastly, the benefits of schedule coordination between the trunk line and feeder lines are discussed in Section 4.5.

## 4.1 Experimental setup

Our models are applicable to any continuous demand functions. For illustrative purposes, we use the following demand functions in our numerical case studies:

$$\lambda_p(x,y) = TrN(x|\mu_{xp}, \sigma_{xp}, 0, L)TrN(y|\mu_{yp}, \sigma_{yp}, 0, W)\Lambda_p \tag{23a}$$
$$\lambda_d(x,y) = TrN(x|\mu_{xd}, \sigma_{xd}, 0, L)TrN(y|\mu_{yd}, \sigma_{yd}, 0, W)\Lambda_d \tag{23b}$$

where $TrN(x|\mu, \sigma, a, b)$ indicates the probability density function of a truncated normal distribution with mean $\mu$, standard deviation $\sigma$, and support $[a, b]$. Note that $\sigma_{xp}$, $\sigma_{yp}$, $\sigma_{xd}$, and $\sigma_{xd}$ can be viewed as proxies for demand heterogeneity. A larger value of $\sigma_{xp}$, $\sigma_{yp}$, $\sigma_{xd}$, or $\sigma_{xd}$ signifies that the spatial distribution of trip origins or destinations is "flatter." Particularly, if $\sigma_{xp} = \sigma_{yp} = \sigma_{xd} = \sigma_{xd} = \infty$, the demand is uniformly distributed over the service region.

Additionally, note that patrons in close proximity to the terminal may prefer to walk rather than take feeder buses. To account for this preference, we assume that patrons with an origin or destination within 300 meters from the terminal, using the Manhattan distance measure, will not take the feeder buses.[7]

Table 2. Parameter values

| Notation | Description | Value | Unit |
|---|---|---|---|
| $L$ | Length of the service region | 3 | km |
| $W$ | Width of the service region | 2 | km |
| $\theta$ | Value of time | 20 | $/h |
| $\pi_s$ | Amortized unit cost per feeder stop infrastructure per operation hour | 0 | $/stop/h |
| $\pi_v$ | Unit cost per bus-km traveled | $0.0314 + 0.0039K$ | $/vehicle·km |
| $\pi_m$ | Unit cost per bus hour | $2.068 + 0.108K + 2\theta$ | $/vehicle·h |
| $\tau_0$ | Bus dwell time per stop | 12/3600 | h/stop |
| $\tau_a$ | Alighting time per patron | 2/3600 | h/patron |
| $\tau_b$ | Boarding time per patron | 4/3600 | h/patron |
| $v_W$ | Walking speed | 2 | km/h |
| $v_I$ | Bus cruise speed | 25 | km/h |
| $t_{f-t}$ | Transfer delay from feeder to trunk transit per patron | 3/60 | h/patron |
| $t_{t-f}$ | Transfer delay from trunk transit to feeder per patron | 3/60 | h/patron |
| $H_{min}$ | Minimum feeder headway | 3/60 | h |
| $H_{max}$ | Maximum feeder headway | 30/60 | h |
| $H_t$ | Headway of the trunk line | 5/60 | h |
| $\varepsilon$ | error tolerance | 0.0001 | – |

Numerical instances in this section use the parameter values displayed in Table 2 (Jaiswal, 2010; Gu et al., 2016; Chen et al., 2017; Mei et al., 2021; Sangveraphunsiri et al., 2022) unless otherwise specified. The service region size ($2 \times 3$ km²) is suitable for representing a quarter of a small town or satellite city connected to a trunk transit system (e.g., the city of Langfang connected to the intercity high-speed rail network, and cities in the San Francisco Bay Area connected to the BART system). The relationships between $\pi_v$, $\pi_m$ and $K$, $\theta$ are obtained using empirical data; see Appendix D for details.

---

[7] Sensitivity analysis shows that our main findings remain valid when this distance threshold varies within a reasonable range. For instance, when the walking zone size increases from 300 m to 500 m, the change in generalized cost savings for our heterogeneous design is less than 0.1%.

Regarding the hyperparameters of the solution algorithm, we find that $m = 30$ and $n = 20$ yield sufficiently accurate numerical solutions. Further increasing $m$ or $n$ renders a relative percentage change in the generalized cost of less than 0.5%. Thus, the values of $m$ and $n$ are set to 30 and 20, respectively. The error tolerance $\varepsilon$ is set to 0.0001. Each numerical instance is solved using various combinations of initial values, consistently converging to the same optimal solution. A typical instance's solution is found within 30 seconds.

## 4.2 Validating the accuracy of the CA cost functions

To examine the accuracy of the CA cost functions, we first convert the optimal CA solution, which comprises continuous functions for headways, line and stop spacings, into a specific feeder system design. This is done using the method described in Appendix C. Following this, we recalculate the generalized cost and each cost component for comparison. We apply a demand pattern with $\Lambda_p = \Lambda_d = 1200$ patrons/h, $\mu_{xp} = \mu_{xd} = 0$, $\mu_{yp} = \mu_{yd} = 0$, $\sigma_{xp} = \sigma_{xd} = L/4$, and $\sigma_{yp} = \sigma_{yd} = W/4$. These parameter values suggest that the demand is concentrated near the terminal, a common scenario where the trunk transit station is located in an area with high neighboring demand. The spatial distribution of the demand is illustrated in Fig. 4. We set the bus capacity at $K = 10$ (e.g., a 10-passenger Ford Transit Van). The converted feeder network, comprising the layout of feeder lines and the locations of stops (denoted by square markers), is illustrated in Fig. 5. The figure reveals that the bus lines and stops are more densely clustered near the terminal, reflecting the demand distribution shown in Fig. 4. It is worth noting that the line and stop locations can be fine-tuned to align with the local street network. For a case study on adapting a conceptual design to a local city street network, please refer to Estrada et al. (2011).

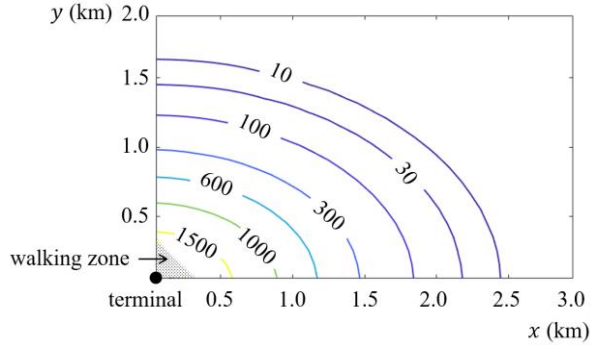

**Fig. 4 The demand densities (patron/km²/h) in the collection direction for a terminal-centered demand**

Percentage errors are computed between the recalculated cost terms and those of the CA cost functions. The results are displayed in Table 3. Note that the error in generalized cost is only 0.73%, and the error in each cost component remains consistently below 2%. This demonstrates the accuracy of our CA cost functions.[8]

---

[8] The error can be further reduced by an improved method for generating the specific network design; see the details in

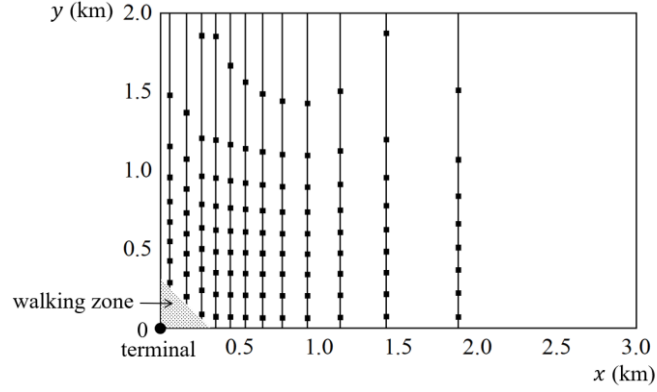

**Fig. 5 The specific feeder system layout**

**Table 3. Percentage errors in cost terms**

| Cost terms | CA | Specific plan | Percentage error |
|---|---|---|---|
| $AC$, h | 98.58 | 97.85 | 0.75% |
| $C_A$, h | 85.96 | 85.38 | 0.68% |
| $C_W$, h | 254.80 | 254.05 | 0.30% |
| $C_T$, h | 125.33 | 123.29 | 1.65% |
| $UC$, h | 466.10 | 462.73 | 0.73% |
| $GC$, h | 564.68 | 560.58 | 0.73% |

## 4.3 Optimal feeder system designs without schedule coordination

We use the same terminal-centered demand pattern as described in Section 4.2 and analyze two scenarios with distinct time values: $\theta = 5$ \$/h for a low-wage city and $\theta = 20$ \$/h for a high-wage city.

First, we examine how the optimal generalized cost varies with bus capacity $K$ in the two scenarios, as depicted in Figs. 6a–b. These figures reveal that the generalized cost is sensitive to $K$ when $K$ is small, but becomes less sensitive for larger $K$ values. The high-wage city exhibits lower sensitivity since the agency cost is associated with a smaller weight. For the same reason, the high-wage city favors a slightly smaller vehicle.[9]

Throughout the remainder of Section 4, we will focus on the optimal feeder designs using the optimal $K$ obtained through exhaustive search.

Figs. 7a and b plot the optimal line spacings and collection-direction headways against the location of lines ($x$) under the two scenarios, respectively. In each scenario, the distributions of spatially-heterogeneous line spacings and headways match the demand distribution (see Fig. 4); that is, greater transit service frequencies and line densities are observed in areas with higher demand densities near the terminal. The high-wage city features higher line densities and service frequencies, as patron costs carry a larger weight. Nonetheless, the discrepancy between low- and high-wage cities remains relatively small. This can be attributed to the fact that 90% of the agency cost, specifically the cost

---

Appendix C.

[9] Actual values of $K$ depend on the available bus models on the market. In addition, the costs of different vehicles might deviate from the unified cost functions derived in Appendix D. Thus, the optimal bus capacity discussed in this study should be viewed as a guideline rather than a definitive solution. Transit agencies are advised to consider vehicles with a size approximate to, but not necessarily equal to, the optimal $K$.

component associated with fleet size, $C_{vh}$, is proportional to the value of time $\theta$; see the formula for $\pi_m$ in Table 2 and its derivation in Appendix D. When $C_{vh}$ is converted to hours by dividing by $\theta$ (see Eq. (11)), the impact of $\theta$ is mostly neutralized. In essence, since the relationship between the value of time and agency cost rates is factored into the model, the generalized cost, as well as the optimal system design, are only marginally influenced by city's wealth level.

Figs. 7c and d display the optimal stop spacings under the two scenarios. As expected, the stop spacing also decreases as the demand density increases and appears to be insensitive to the value of time. In both scenarios, *the optimal stop spacing varies in a rather wide range*, highlighting the need for heterogeneous stop spacing planning under the spatially heterogeneous demand.

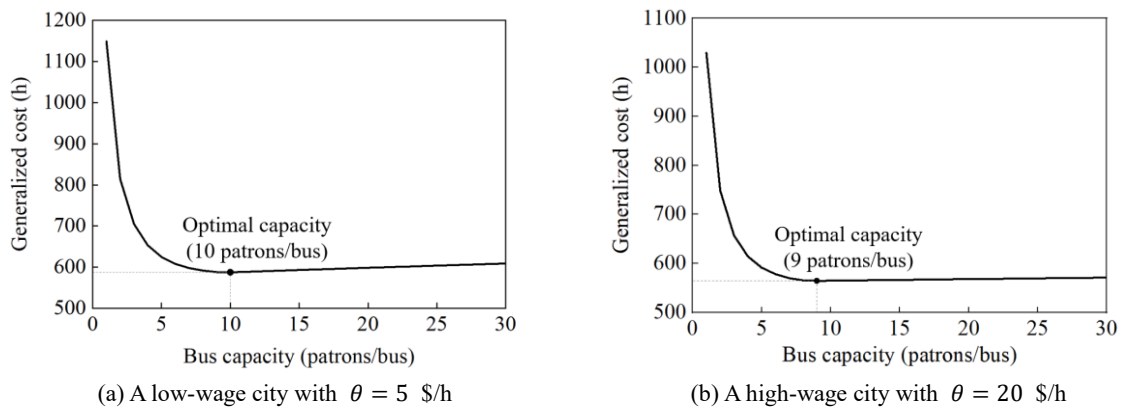

(a) A low-wage city with $\theta = 5$ \$/h  (b) A high-wage city with $\theta = 20$ \$/h

**Fig. 6 Generalized cost versus bus capacity**

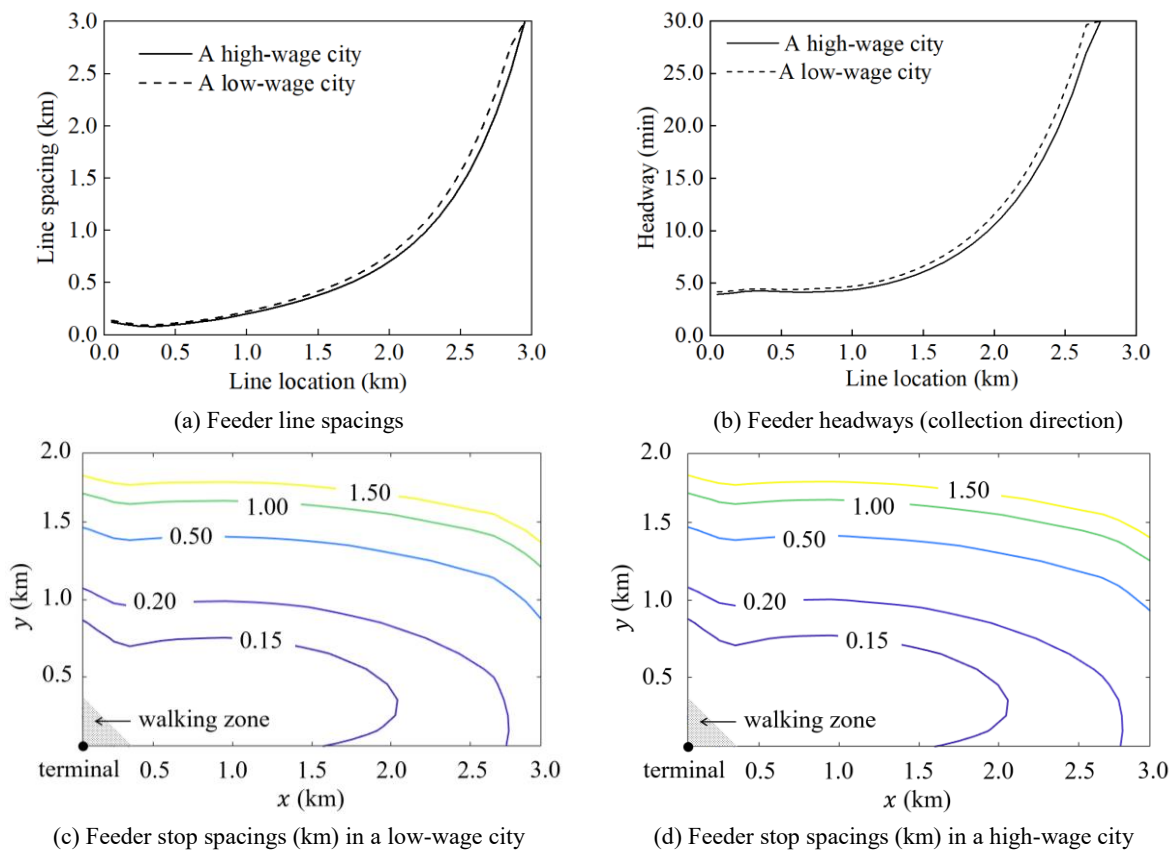

(a) Feeder line spacings  (b) Feeder headways (collection direction)

(c) Feeder stop spacings (km) in a low-wage city  (d) Feeder stop spacings (km) in a high-wage city

**Fig. 7 Optimal design variables under the terminal-centered demand**

Next, we compare our heterogeneous design against three benchmark designs, including: (i) an optimal design with heterogeneous line spacings and headways but uniform stop spacing; (ii) an optimal design featuring spatially-heterogeneous headways with uniform line and stop spacings; and (iii) an optimal design with a uniform line spacing, stop spacing, and headway. All four optimal designs use the same demand pattern and a value of time set at 20 $/h. Results are presented in Table 4, where the range of optimal values for each heterogeneous decision variable is shown in brackets, followed by the average value in parentheses.

Table 4. Optimal heterogeneous, partially-homogeneous, and homogeneous designs under terminal-centered demand

| Variable | Heterogeneous design | Uniform stop spacing | Uniform line and stop spacings | Uniform line and stop spacings, and headway |
|---|---|---|---|---|
| $K^*$ | 9 | 9 | 12 | 22 |
| $B^*(x,y)$, km | [0.12, 2.00] (0.23) | 0.16 | 0.15 | 0.16 |
| $S_l^*(x)$, km | [0.08, 3.00] (0.24) | [0.09, 3.00] (0.25) | 0.20 | 0.22 |
| $H_{lp}^*(x)$, min | [3.9, 30.0] (10.7) | [4.0, 30.0] (11.0) | [3.0, 30.0] (14.2) | 5.2 |
| $H_{ld}^*(x)$, min | [5.0, 30.0] (11.1) | [5.0, 30.0] (11.3) | [5.0, 30.0] (14.8) | 5.2 |
| $AC$, h | 98.51 | 102.02 | 94.23 | 139.89 |
| $UC$, h | 465.89 | 470.19 | 487.80 | 507.19 |
| $GC$, h | 564.39 | 572.20 | 582.04 | 647.09 |
| $GC$ saving | **1.36%** (compared to design with uniform stop spacing) | **1.68%** (compared to design with uniform line and stop spacings) | **10.05%** (compared to the homogeneous design) | – |

The last row of Table 4 demonstrates that, compared to the homogeneous design, heterogenizing the headway results in a cost saving of 10.05%. Additionally, further heterogenizing the line spacing yields an extra saving of 1.68%, while heterogenizing the stop spacing contributes an added saving of 1.36%. Note that each number is computed by comparing the $GC$ against a "more homogeneous" design, which is the one in the column immediately to the right of the current design in Table 4. The last number is a conservative estimate, as it is derived from a design with optimized (though uniform) stop spacing. Should stop spacing be disregarded or assumed constant in the model, as in previous studies such as Chang and Schonfeld (1991) and Yang et al. (2020), even greater savings would be realized. For example, compared to using a fixed stop spacing of 500 m, the cost saving stemmed from our heterogeneous design amounts to 11.37%. Furthermore, larger savings are also observed under alternative demand patterns, as will be discussed shortly.

Lastly, we compare the generalized cost of the heterogeneous design to the cost of the optimal design without accounting for the impacts of boarding and alighting passengers on dwell times and transfer times (details are omitted in the interest of brevity). In the latter design, the bus dwell time is set at 30 seconds, as in Gu et al. (2016). Numerical results across various demand levels reveal that the relative percentage error lies between 2–5%. This underscores the importance of incorporating passenger boarding and alighting processes in the model.

### 4.4 Sensitivity analyses

In this section, we examine the sensitivity of the optimal heterogeneous design and cost savings in comparison to the benchmark designs. Sensitivity analyses concerning the demand patterns, demand rates, service region sizes, and feeder line layouts are furnished in subsections 4.4.1–4.4.4, respectively.

*4.4.1 Sensitivity to the demand pattern*

We investigate the benefit of heterogenous design under two distinct demand patterns for a high-wage city ($\theta = 20$ \$/h). The first pattern remains terminal-centered but exhibits a more concentrated spatial distribution with $\sigma_{xp} = \sigma_{xd} = L/8$ and $\sigma_{yp} = \sigma_{yd} = W/8$. The other parameter values are consistent with those in Section 4.3. The outcomes of the optimal heterogeneous design and the three benchmark designs are presented in Table 5. The last row of this table is derived in a similar fashion to Table 4.

Table 5. Optimal designs when demand is more concentrated near the terminal

| Variable | Heterogeneous design | Uniform stop spacing | Uniform line and stop spacing | Uniform line and stop spacing, and headway |
|---|---|---|---|---|
| $K$, pax | 8 | 13 | 13 | 33 |
| $B^*(x,y)$, km | [0.09, 2.00] (0.34) | 0.13 | 0.13 | 0.14 |
| $S_l(x)$, km | [0.03, 3.00] (0.23) | [0.04, 3.00] (0.26) | 0.13 | 0.20 |
| $H_{lp}(x)$, min | [3.9, 30.0] (19.4) | [4.1, 30.0] (19.7) | [3.0, 30.0] (21.6) | 5.9 |
| $H_{ld}(x)$, min | [5.0, 30.0] (19.6) | [5.0, 30.0] (19.8) | [5.0, 30.0] (21.9) | 5.9 |
| $AC$, h | 83.42 | 88.21 | 96.09 | 142.72 |
| $UC$, h | 313.50 | 322.88 | 330.84 | 384.98 |
| $GC$, h | 396.92 | 411.09 | 426.93 | 527.70 |
| $GC$ gap | **3.45%** | **3.71%** | **19.1%** | – |

Table 5 shows significantly larger percentage cost savings under the more concentrated demand pattern compared to Table 4. For example, heterogenizing the stop spacings now produces a 3.45% cost saving. This suggests that the *heterogeneous design is more advantageous when demand is more spatially heterogeneous*. Also, designs with higher degrees of heterogeneity tend to favor smaller feeder buses, as the optimal spatially-heterogeneous line and stop spacings align more closely with the heterogeneous demand (see Fig. 7), *rendering the patrons more evenly distributed across the lines and vehicles*.

The second demand pattern employs the same standard deviation parameters as in Section 4.3, but with demand centered at the far end of the service region, i.e., $\mu_{xp} = \mu_{xd} = L$ and $\mu_{yp} = \mu_{yd} = W$. This remotely-centered pattern also has real-world examples.[10] Other parameters remain the same as before. The results under this demand pattern are provided in Table 6.

Table 6 demonstrates that merely heterogenizing stop spacings results in an even greater cost saving of 4.17%. This is likely due to the better alignment of optimal stop spacings with this demand pattern. Specifically, $B^*(x,y)$ is small near the service region's top edge due to high demand density and low onboard patron flow, while it is large near the bottom owing to low demand density and high onboard patron flow; see Eq. (20g). Recall that the advantages of our design would be even more pronounced when compared to previous feeder optimization models that utilize a constant, non-optimized stop spacing.

---

[10] For instance, the Palm Springs and Fairview Park in Hong Kong are large-scale residential estates housing most of the population in the surrounding region. However, these estates are situated several kilometers away from the nearest metro station.

Table 6. Optimal designs when demand is centered at the remote end

| Variable | Heterogeneous design | Uniform stop spacing | Uniform line and stop spacing | Uniform line and stop spacing, and headway |
|---|---|---|---|---|
| $K$, pax | 11 | 11 | 12 | 22 |
| $B^*(x,y)$, km | [0.08, 2.00] (0.28) | 0.24 | 0.24 | 0.25 |
| $S_l(x)$, km | [0.10, 2.29] (0.19) | [0.10, 2.41] (0.19) | 0.18 | 0.20 |
| $H_{lp}(x)$, min | [4.4, 30.0] (10.1) | [4.5, 30.0] (10.2) | [3.1, 30.0] (14.1) | 5.2 |
| $H_{ld}(x)$, min | [5.0, 30.0] (10.3) | [5.0, 30.0] (10.5) | [5.0, 30.0] (14.6) | 5.3 |
| $AC$, h | 122.23 | 122.72 | 117.34 | 145.66 |
| $UC$, h | 813.40 | 853.58 | 862.96 | 877.85 |
| $GC$, h | 935.62 | 976.30 | 980.30 | 1023.50 |
| $GC$ gap | **4.17%** | **0.41%** | **4.22%** | – |

### 4.4.2 Sensitivity to the demand rate

In this subsection, we examine the terminal-centered demand pattern with $\Lambda_p = \Lambda_d$ taking a smaller value (400 patrons/h). The results are provided in Table 7. Comparing the cost savings with those in Table 4 reveals that the percentage cost savings of the optimal heterogeneous design increase moderately as demand diminishes. In addition, lower demand entails fewer lines served by smaller vehicles operating at lower frequencies, which is a logical outcome. However, the stop spacings remain largely unaffected by the demand rate.

### 4.4.3 Sensitivity to the service region's size

In this analysis, we continue to use the terminal-centered demand pattern but examine a smaller service region with $L = 1.5$ km and $W = 1$ km. To ensure a fair comparison, we keep the average demand density (i.e., $\Lambda_p/LW$ and $\Lambda_d/LW$) consistent with the instance in Table 4. The results are summarized in Table 8. The table indicates that the percentage cost savings of the heterogeneous design increase as the service region shrinks, especially the cost saving resulting from spatially-heterogenizing headways. This time, the headways and stop spacings demonstrate relative insensitivity to changes in service region size, while the average line spacing increases significantly as the service region contracts. The optimal $K$ also decreases, since the demand carried by each line is lower.

Table 7. Optimal designs under terminal-centered demand with lower rates

| Variable | Heterogeneous design | Uniform stop spacing | Uniform line and stop spacing | Uniform line and stop spacing, and headway |
|---|---|---|---|---|
| $K$, pax | 7 | 7 | 8 | 18 |
| $B^*(x,y)$, km | [0.13, 2.00] (0.25) | 0.17 | 0.16 | 0.17 |
| $S_l(x)$, km | **[0.22, 3.00] (0.49)** | [0.22, 3.00] (0.49) | 0.40 | 0.42 |
| $H_{lp}(x)$, min | **[4.4, 30.0] (13.3)** | [4.6, 30.0] (13.5) | [3.2, 30.0] (15.4) | 6.7 |
| $H_{ld}(x)$, min | **[5.0, 30.0] (13.5)** | [5.0, 30.0] (13.7) | [5.0, 30.0] (15.8) | 6.7 |
| $AC$, h | 43.77 | 40.62 | 40.10 | 54.89 |
| $UC$, h | 173.70 | 175.65 | 182.58 | 195.01 |
| $GC$, h | 212.74 | 216.27 | 222.68 | 249.90 |
| $GC$ gap | **1.63%** | **2.88%** | **10.89%** | – |

The results in Tables 4–8 reveal that spatially-heterogenizing headways always yields the largest benefit. Nevertheless, *heterogenizing stop spacings also contributes to moderate yet consistent improvements*.

Table 8. Optimal designs with a $1 \times 1.5$ km² service region

| Variable | Heterogeneous design | Uniform stop spacing | Uniform line and stop spacing | Uniform line and stop spacing, and headway |
|---|---|---|---|---|
| $K$, pax | 6 | 6 | 9 | 12 |
| $B^*(x,y)$, km | [0.10, 1.00] (0.20) | 0.12 | 0.11 | 0.13 |
| $S_I(x)$, km | **[0.17, 1.50] (0.35)** | [0.17, 1.50] (0.36) | 0.37 | 0.29 |
| $H_{Ip}(x)$, min | [3.6, 30.0] (10.7) | [3.9, 30.0] (11.0) | [3.0, 30.0] (12.8) | 4.4 |
| $H_{Id}(x)$, min | [5.0, 30.0] (11.2) | [5.0, 30.0] (11.4) | [5.0, 30.0] (13.6) | 5.8 |
| $AC$, h | 17.78 | 18.90 | 15.68 | 28.18 |
| $UC$, h | 85.23 | 86.35 | 93.91 | 108.09 |
| $GC$, h | 103.01 | 105.25 | 109.60 | 136.27 |
| $GC$ gap | **2.13%** | 3.97% | 19.57% | – |

*4.4.4 Sensitivity to the feeder line layout*

In this subsection, we compare two alternative feeder line layouts: (i) feeder buses travel along the $y$-direction for patron pick-up or drop-off and move along the $x$-direction to and from the terminal without stopping; and (ii) feeder buses load and unload passengers along the $x$-direction and perform nonstop travel along the $y$-direction. We let the rectangular service region's aspect ratio, $\frac{W}{L}$, vary from 0.1 to 2, and plot the generalized costs per patron for the two layouts in Fig. 8. The service region's area $L \times W$ and other parameter values are kept consistent with those in Section 4.3 (with $\theta = 20$ \$/h). The figure reveals that the generalized cost is consistently lower *when buses make stops along the shorter side of the service region*. In this configuration, more bus lines are deployed along the longer side, resulting in each bus carrying fewer patrons and reduced boarding delays. The cost gap between the two layouts is substantial when the service region is "slim." For example, with $\frac{W}{L} = 2$ or 0.5, the more efficient layout outperforms the alternative by 6.49% in generalized cost. This finding highlights the importance of selecting the appropriate layout, particularly when the service region is more oblong than square. Notably, this straightforward result has not been previously documented in the literature; for example, Sivakumaran et al. (2012) arranged the feeder lines along the longer side of the service region, which our analysis indicates is a suboptimal design.

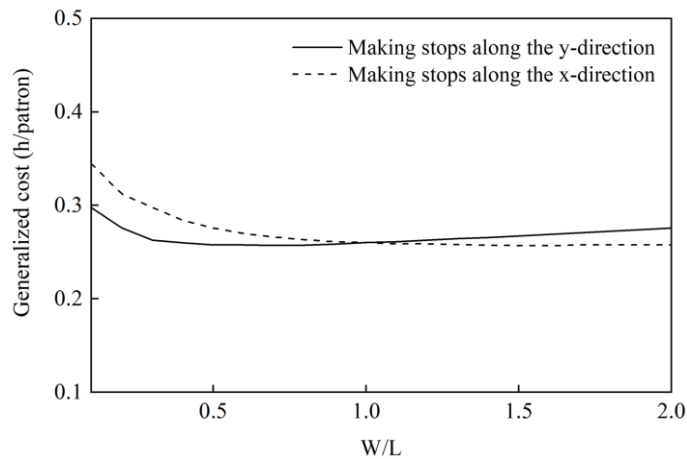

**Fig. 8 Comparison between two alternative feeder line layouts**

**4.5 Benefits of schedule coordination**

We now explore the benefits of schedule coordination. The feeder network is reoptimized considering schedule coordination in the collection direction only, in the distribution direction only, and in both directions simultaneously. The percentage of generalized cost savings compared to the uncoordinated case is plotted against $H_t \in \{5, 7.5, 10\}$ minutes in Fig. 9. Other parameters remain the same as those in Section 4.3 (with $\theta = 20\$/h$).

Fig. 9 demonstrates the sizeable benefits of schedule coordination. Coordination in the collection direction brings cost savings ranging from 7.42% to 10.95%. Coordination in the distribution direction yields a smaller but still noteworthy benefit (-0.04% to 7.35%), particularly when $H_t$ is large. Moreover, coordinating in both directions achieves savings of up to 20.10%. Intriguingly, this value slightly exceeds the combined percentage cost savings of coordinating in individual directions. This can be attributed to two possible reasons: (i) waiting time constitutes a large portion of the generalized cost; and (ii) simultaneously coordinating the schedules in both directions may facilitate more efficient optimization of line spacings, stop spacings, and the vehicle size (factors that concurrently impact the generalized costs in both directions).

Additionally, the figure indicates that coordination becomes increasingly advantageous as trunk-line headway grows. This is owing to two reasons: (i) the waiting time saved by coordination rises with trunk-line headway (see Section 2.5); and (ii) the agency cost savings also grow as larger feeder bus headways (in multiples of $H_t$) are enforced.

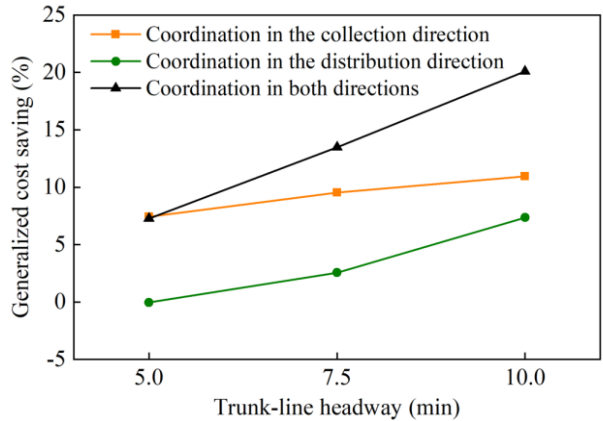

**Fig. 9 Generalized cost saving of schedule coordination**

## 5. Conclusions

We propose continuous approximation models for designing an optimal heterogeneous feeder bus network catering to spatially heterogeneous demand to and from a rail terminal. These models identify the optimal feeder service headways, line spacings, stop spacings, and vehicle size that minimize the generalized system cost. An iterative method was developed to find the optimal solution, leveraging the analytical properties of the optimal solution. Extensive numerical analyses were conducted to validate the accuracy of the CA models, showcase the effectiveness of our solution method, and unveil new insights that have practical implications.

Our models stand out by considering the effects of passenger boarding and alighting on bus dwell times and transfer times, while optimizing heterogeneous stop spacings alongside line spacings and headways. Numerical results show that more accurately modeling dwell and transfer times yields a 2–5% difference in generalized cost compared to prior models with constant dwell times. Furthermore, optimizing heterogeneous stop spacings results in additional cost savings up to 4% compared to designs with optimized uniform stop spacing, and the savings could be over 10% when contrasted with previous studies that only optimized line spacings and headways. Additionally, our model that integrates schedule coordination reveals greater benefits (cost savings up to 20%) when trunk and feeder services are coordinated in both feeder travel directions, potentially surpassing the combined savings of coordination in each individual direction. This suggests that trunk and feeder schedules should be coordinated in both service directions whenever feasible.

Other important findings and their practical implications are summarized as follows:

(i) The benefit of the heterogeneous design largely depends on the demand's spatial pattern. Greater benefits were observed for more heterogeneous demand and when the spatial concentration of demand shifted away from the rail terminal. Higher demand rates and larger service region sizes tend to reduce the benefit.

(ii) Compared to homogeneous or partially-homogeneous feeder network designs, an optimal heterogeneous feeder network uses smaller buses. This is because the heterogeneous design better accommodates heterogeneous demand, resulting in more evenly distributed passenger loading across lines and vehicles.

(iii) Feeder lines should be designed to enable buses to travel along the shorter side of the service region for pick-up and drop-off, while traveling nonstop along the longer side to the rail station. Selecting the appropriate layout can further reduce system costs by 6% or more, depending on the service region's shape.

(iv) Our models and algorithms (including the algorithm for generating a specific feeder system design) can be readily used by local transit agencies to design a fixed-route feeder network within a rectangular service region. The generated plan can be fine-tuned to align with the local street network.

Our models can be moderately adapted to include currently overlooked minor cost items, such as those related to bus deadheading trips, layover times, and environmental impacts (Cheng et al., 2016). Additionally, we are currently exploring the comparison between the optimal fixed-route feeder system and alternative trunk-line access modes, including shared bikes, modular vehicles, ride-hailing, and flex-route feeders. Built upon our models, we aim to further optimize a heterogeneous trunk-and-feeder transit network that serves a city-wide demand. This research direction is also under investigation.

## Acknowledgments:

This study is supported by a General Research Fund (Project No. 15224818) provided by the Research Grants Council of Hong Kong.

## Appendix A. Table of notations

Table A1. List of notations

| Notation | Description | Unit |
|---|---|---|
| *Decision variables* | | |
| $B(x, y)$ | Stop spacing at location $(x, y)$ | km |
| $S_l(x)$ | Line spacing at location $x$ | km |
| $H_{Ip}(x)$ | Feeder bus headway in the patron-collection direction at location $x$ | h |
| $H_{Id}(x)$ | Feeder bus headway in the patron-distribution direction at location $x$ | h |
| $K$ | The patron-carrying capacity of a feeder bus | patrons/vehicle |
| *Cost terms and parameters (all cost terms are per hour of operations)* | | |
| $AC$ | Total agency cost | h |
| $UC$ | Total user cost | h |
| $GC$ | Generalized cost | h |
| $C_A$ | Total access and egress time | h |
| $C_W$ | Total wait and transfer time | h |
| $C_{Wp}$ | Wait and transfer time in the collection direction | h |
| $C_{Wd}$ | Wait and transfer time in the distribution direction | h |
| $C_T$ | Total in-vehicle travel time | h |
| $C_{T1}$ | In-vehicle travel time to overcome the distance | h |
| $C_{T2}$ | Patrons' time loss due to bus deceleration, acceleration, door opening and closing | h |
| $C_{T3}$ | Patrons' time loss due to boarding and alighting behavior | h |
| $C_s$ | Stop infrastructure cost | h |
| $C_{vk}$ | Operating cost associated with bus-km traveled | h |
| $C_{vh}$ | Operating cost related to bus fleet size | h |
| $V_h$ | Bus fleet size | |
| $V_{h1}$ | Number of buses that are actively travelling between stops | |
| $V_{h2}$ | Number of buses needed due to deceleration, acceleration, door opening and closing at stops | |
| $V_{h3}$ | Number of buses required for passenger boarding and alighting | |
| $\pi_s$ | The amortized unit cost of feeder stop infrastructure per operation hour | \$/stop/h |
| $\pi_v$ | The unit cost per bus-km traveled | \$/vehicle·km |
| $\pi_m$ | The unit cost per bus hour | \$/vehicle·h |
| *Other parameters and variables* | | |
| $\lambda_p(x,y)$ | Demand density in the patron-collection direction at location $(x,y)$ | patrons/km²/h |
| $\lambda_d(x,y)$ | Demand density in the patron-distribution direction at location $(x,y)$ | patrons/km²/h |
| $\Lambda_p$ | Total demand in the patron-collection direction | patrons/h |
| $\Lambda_d$ | Total demand in the patron-distribution direction | patrons/h |
| $\Lambda_p{}'$ | Total demand in the collection direction after removing walking patrons | patrons/h |
| $\Lambda_d{}'$ | Total demand in the distribution direction after removing walking patrons | patrons/h |
| $\theta$ | Value of time | \$/h |
| $W$ | Width of the service region | km |
| $L$ | Length of the service region | km |
| $v_W$ | Walking speed | km/h |
| $v_I$ | Feeder bus cruise speed | km/h |
| $\tau_0$ | Time loss per stop due to bus deceleration, acceleration, door opening and closing | h/stop |
| $\tau_a$ | Alighting time per patron | h/patron |
| $\tau_b$ | Boarding time per patron | h/patron |
| $t_{f-t}$ | Transfer delay from feeder stop to the trunk-line platform per patron | h/patron |
| $t_{f-t}$ | Transfer delay from trunk-line platform to feeder stop per patron | h/patron |
| $H_t$ | Headway of the trunk line | h |
| $H_{min}$ | Minimum headway of a feeder line | h |
| $H_{max}$ | Maximum headway of a feeder line | h |

## Appendix B. Derivation of the optimality properties

Combining all the cost metrics, we can rewrite the generalized cost (15a) as:

$$GC = \left\{\frac{\pi_m}{\mu}(\tau_a+\tau_b)\Lambda + \int_{x=0}^{L}\left(\frac{x}{v_I}(\Lambda_{px}(x)+\Lambda_{dx}(x)) + \left(t_{f-t}+\frac{H_t}{2}\right)\Lambda_{px}(x) + t_{t-f}\Lambda_{dx}(x)\right)dx + \right.$$

$$\int_{y=0}^{W}\frac{y}{v_I}\left(\Lambda_{py}(y)+\Lambda_{dy}(y)\right)dy\right\} + \left\{\int_{x=0}^{L}\left[\frac{\pi_v(W+x)}{\theta S_I(x)}\left(\frac{1}{H_{Ip}(x)}+\frac{1}{H_{Id}(x)}\right) + \frac{\pi_m(W+x)}{\theta S_I(x)v_I}\left(\frac{1}{H_{Ip}(x)}+\frac{1}{H_{Id}(x)}\right) + \right.\right.$$

$$\frac{S_I(x)}{4v_W}(\Lambda_{px}(x)+\Lambda_{dx}(x)) + \frac{H_{Ip}(x)}{2}\Lambda_{px}(x) + \frac{H_{Id}(x)}{2}\Lambda_{dx}(x) + \frac{\tau_a}{2}S_I(x)H_{Ip}(x)[\Lambda_{px}(x)]^2 +$$

$$\left.\frac{\tau_b}{2}S_I(x)H_t[\Lambda_{dx}(x)]^2 + \tau_bS_I(x)H_{Ip}(x)M_{px}(x) + \tau_aS_I(x)H_{Id}(x)M_{dx}(x)\right]dx\right\} +$$

$$\left\{\int_{x=0}^{L}\int_{y=0}^{W}\left[\frac{\pi_s}{\theta}\frac{1}{S_I(x)B(x,y)} + \frac{\pi_m}{\theta}\frac{\tau_0}{S_I(x)B(x,y)}\left(\frac{1}{H_{Ip}(x)}+\frac{1}{H_{Id}(x)}\right) + \frac{B(x,y)}{4v_W}[\lambda_p(x,y)+\lambda_d(x,y)] + \right.\right.$$

$$\left.\left.\frac{\tau_0\Lambda_{pxy}(x,y)}{B(x,y)} + \frac{\tau_0\Lambda_{dxy}(x,y)}{B(x,y)}\right]dydx\right\} \quad (B1)$$

The part of the objective function in the first pair of braces is constant. The part in the second pair of braces contains single integrals relating to either $x$ or $y$ (involving decision functions $S_I(x)$, $H_{Ip}(x)$, $H_{Id}(x)$). Meanwhile, the part in the third pair of braces contains double integrals related to both $x$ and $y$ (involving the decision function $B(x,y)$).

Note that (B1) is a convex function when only one decision variable ($S_I(x)$, $H_{Ip}(x)$, $H_{Id}(x)$, or $B(x,y)$) is considered, and the other variables are fixed. (This type of functions is known as the *coordinate-wise convex functions*, which are not necessarily convex themselves.) Further note that constraints (15c)–(15f) act as boundary constraints when only one decision variable is considered. As a result, the optimal solution for each decision variable can be developed analytically from first-order derivatives, assuming the other variables are fixed. These optimal solutions are:

$$H_{Ip}^*(x) = mid\left\{H_{min}, \min\left\{\frac{K}{\Lambda_{px}(x)S_I^*(x)}, H_{max}\right\}, \sqrt{\frac{2\alpha(x)\left(S_I^*(x)\right)^{-1}}{\Lambda_{px}(x)+\beta_p(x)S_I^*(x)}}\right\} \quad (B2)$$

$$H_{Id}^*(x) = mid\left\{\max\{H_{min},H_t\}, \min\left\{\frac{K}{\Lambda_{dx}(x)S_I^*(x)}, H_{max}\right\}, \sqrt{\frac{2\alpha(x)\left(S_I^*(x)\right)^{-1}}{\Lambda_{dx}(x)+\beta_d(x)S_I^*(x)}}\right\} \quad (B3)$$

$$S_I^*(x) = \min\left\{\sqrt{\frac{2\alpha(x)\left(\left(H_{Ip}^*(x)\right)^{-1}+\left(H_{Id}^*(x)\right)^{-1}\right)+2\gamma(x)}{\frac{\Lambda_{px}(x)+\Lambda_{dx}(x)}{2v_W}+\tau_b[\Lambda_{dx}(x)]^2 H_t+\beta_p(x)H_{Ip}^*(x)+\beta_d(x)H_{Id}^*(x)}}, \frac{K}{\Lambda_{px}(x)H_{Ip}^*(x)}, \frac{K}{\Lambda_{dx}(x)H_{Id}^*(x)}\right\} \quad (B4)$$

$$B^*(x,y) = \sqrt{\frac{4v_W\left[\frac{1}{\mu S_I^*(x)}\left(\pi_s+\frac{\pi_m\tau_0}{H_{Ip}^*(x)}+\frac{\pi_m\tau_0}{H_{Id}^*(x)}\right)+\tau_0\left(\Lambda_{pxy}(x,y)+\Lambda_{dxy}(x,y)\right)\right]}{\lambda_p(x,y)+\lambda_d(x,y)}} \quad (B5)$$

where function $mid\{z_1, z_2, z_2\}$ returns the middle one of the three arguments.

## Appendix C. Generating a specific feeder system design

Given $S_I^*(x_i)$ and $B^*(x_i, y_j)$ ($i = 1,2,...,n$; $j = 1,2,...,m$), we use the following three steps to generate line and stop locations. Step 1 generates line locations that closely match $S_I^*(x_i)$. Step 2 determines the continuous stop spacing function for each line generated in Step 1. Step 3 generates the stop locations for each line.

Step 1. Apply the spline curve fitting method to fit a cubic spline function $S_I(x)$ to $S_I^*(x_i)$ ($i = 1,2,...,n$). Next, place one line at every $x$ where $\int_{z=0}^{x}\frac{dz}{S_I(x)} = k + 0.5, k = 1,2,...$. Here, 0.5 is added to the RHS to ensure the first line's catchment zone resides on both sides of that line. The resulting line location set is denoted by $\Omega_S = \{x_p : p = 1,2,...,N^S\}$, where $N^S$ is the number of lines, and $x_p^S$ the location of the $p$-th line satisfying $0 < x_1 < x_2 < \cdots < x_{N^S} < L$.

Step 2. Use the gridded data interpolation method to fit a piecewise cubic function $B(x,y)$ to $B^*(x_i, y_j)$ ($i = 1,2,...,n; j = 1,2,...,m$). The continuous stop spacing function on the $p$-th line is $B(x_p, y)$.

Step 3. For each line $p \in \{1,2,...,N^S\}$, place one stop at every $y$ where $\int_{z=0}^{y}\frac{dz}{B(x_p,y)} = k + 0.5$, $k = 1,2,...$. The resulting stop location set is denoted by $\Omega_p^B = \{(x_p, y_q): q = 1,2,...,N_p^B\}$, where $N_p^B$ is the number of stops on the $p$-th line. Coordinates $y_q$ satisfy $0 < y_1 < y_2 < \cdots < y_{N_p^B} < W$.

Finally, the headways of each line are obtained from $H_{Ip}^*(x)$ and $H_{Id}^*(x)$ at $x_p \in \Omega_S$.

Note that the CA solution renders $\int_{x=0}^{L}\frac{1}{S_I^*(x)}dx$ feeder lines, and this number is not always close to an integer. The above method guarantees finding the nearest integer number of lines. However, when that number is small, the percentage error could be relatively large (capped by $0.5\left(\int_{x=0}^{L}\frac{1}{S_I^*(x)}dx\right)^{-1}$). To further reduce this error and improve the converted specific design, we can evenly allocate the fractional residual number of lines (positive or negative) to all lines by proportionally increasing or decreasing the line spacings. Similarly, stop spacings on each line can also be adjusted to eliminate the fractional residuals. In addition, one can reoptimize stop spacings and headways using the CA model after line locations are determined by Step 1 of the above method. This will make the specific design more efficient. For simplicity, these fine-tuning steps are not included in our numerical case studies.

# Appendix D. Transit agency cost rates

The unit operating cost per bus-hour, $\pi_m$, is primarily determined by the amortized vehicle price, maintenance costs, and staff wages. The vehicle purchase and maintenance costs can be assumed to be an affine function of bus capacity (Oldfield and Bly, 1985), while the staff wages are proportional to the value of time. In other words, we can express $\pi_m$ as:

$$\pi_m = a_m + b_m K + c_m \theta \tag{D1}$$

where $a_m$, $b_m$, and $c_m$ are the cost coefficients.

The unit operating cost per bus-km traveled, $\pi_v$, is mainly determined by the fuel costs, which are related to $K$ but not to $\theta$. Thus, we can assume an affine relationship between $\pi_v$ and $K$ as follows:

$$\pi_v = a_v + b_v K \tag{D2}$$

where $a_v$ and $b_v$ are the cost coefficients.

To calibrate the cost coefficients, we utilized data from two popular bus models: the Volvo 8900 for a full-sized bus and the Ford Transit for a medium-sized one. Table D1 displays the capacities, purchase costs, fuel consumption rates, and service lives of these vehicle models.

We assume two employees are assigned to each bus during operation hours, including one driver and one manpower for maintenance, management, administration, and support services combined. Therefore, we set $c_m = 2$. To amortize the bus purchase costs, we assume that each bus operates for 16 hours a day. Additionally, we inflate the bus purchase cost by 50% to account for other cost items associated with the bus fleet, such as maintenance, insurance, and facility rent. Based on these assumptions, we determine that $a_m = 2.068\$/\text{vehicle}\cdot\text{h}$, and $b_m = 0.108\$/\text{h}$.

Table D1. Data of Volvo 8900 and Ford Transit[11]

| Models | Capacity | Price ($) | Fuel consumption rate (liter/100 km) | Service life (year) |
|---|---|---|---|---|
| Volvo 8900 | 80 | 500,000 | 26 | 12 |
| Ford Transit | 17 | 76,000 | 7.4 | 5 |

For the bus-km-related cost coefficients, we assume a prevailing gasoline price of 1.1\$/liter[12]. The bus-km-related costs are also inflated by 20% to cover the cost items other than fuel, such as mileage-related depreciation, vehicle deterioration, and toll. Based on these assumptions, we determine that $a_v = 0.0314\$/\text{vehicle}\cdot\text{km}$, and $b_v = 0.0039\$/\text{km}$.

---

[11] The passenger-carrying capacity of a full-sized (single-deck, not articulated) bus like the Volvo 8900 may vary depending on the seat plan requested of the transit agency. For our analysis, we assume a capacity of 80 passengers, including standees. The purchase cost of Volvo 8900 was obtained from Horrox and Casale (2019). Its fuel consumption rate was extracted from the Volvo Buses Environmental Blog (https://volvobusesenvironmentblog.wordpress.com/). As for the Ford Transit, we obtained its seat capacity and purchase cost from the Ford Transit price list (https://www.ford.co.uk/content/dam/guxeu/uk/documents/price-list/commercial-vehicles/PL-Transit_Minibus.pdf), while its fuel consumption rate came from the Ford Media Center (https://media.ford.com/content/fordmedia/feu/nl/nl/news/2014/05/15/ford-launches-first-18-seat-transit-minibus--offers-improved-eff.html). We learned the service lives of both models from Laver et al. (2007).

[12] https://www.eia.gov/petroleum/gasdiesel/, accessed on June 15, 2022.